\documentclass[a4paper]{mem}
\usepackage{natbib}
\usepackage{graphicx}
\usepackage[a4paper]{hyperref}
\idline{73}{23}
\def\gtsima{$\; \buildrel > \over \sim \;$}
\def\ltsima{$\; \buildrel < \over \sim \;$}
\def\prosima{$\; \buildrel \propto \over \sim \;$}
\def\gsim{\lower.5ex\hbox{\gtsima}}
\def\lsim{\lower.5ex\hbox{\ltsima}}
\def\simgt{\lower.5ex\hbox{\gtsima}}
\def\simlt{\lower.5ex\hbox{\ltsima}}
\def\simpr{\lower.5ex\hbox{\prosima}}

\def\eg{{\frenchspacing e.g. }}

\def\hinv{$h^{-1}$}

\def\vac#1{}

\newcommand{\gr}{$\gamma$-ray }
\newcommand{\epm}{e$^\pm$ }

\newcommand{\pnd}{$\pi^0$-decay }

\newcommand{\ic}{IC }
\begin{document}
   \title{Non-thermal particles in the intergalactic and intracluster medium 
}
   \author{F. Miniati 
}

   \institute{
Max-Planck-Institut f\"ur Astrophysik,
Karl-Schwarzschild-Str. 1, 85740, Garching, Germany 
             }

   \abstract{ I present a review of nonthermal processes in the large
   scale structure of the universe.  After examining the properties of
   cosmological shock waves and their role as particle accelerators, I
   discuss the main observational facts, from radio to \gr and describe 
   the processes that are thought be responsible for the observed
   nonthermal emissions. Finally, I emphasize the important role of
   $\gamma$-ray astronomy for the progress in the field. Non
   detections (upper limits) at these photon energies have already
   allowed us to draw important conclusions. Future observations will
   tell us more about the physical conditions in the intracluster
   medium, physics of shocks dissipation, aspects of CR acceleration.

   \keywords{
acceleration of particles  ---  cosmology: large-scale structure
of universe  ---  galaxies: clusters: general  ---
gamma rays: theory  ---  radiation mechanism: non-thermal  ---   shock waves

               }
   }
   \authorrunning{F. Miniati}
   \titlerunning{Non-thermal particles in the IGM}
   \maketitle
\section{Introduction} \label{intro.se}
The existence of extended regions populated by cosmic ray electrons 
(CRes) in at least some clusters of galaxies has been apparent since the
discovery of diffuse, non-thermal radio 
emissions from the Perseus and Coma clusters more than thirty years ago
 \citep{leel61,willson70}.
Their importance as indicators of physical processes in the
cluster media has grown in recent years as the
number of detected diffuse radio emission clusters has increased, as reports
have appeared of possible diffuse non-thermal emissions in the 
hard X-ray (HXR) and extreme ultra-violet (EUV) 
bands \citep[\eg][]{lieuetal96a,fufeetal99} and as the evidence has 
mounted for a rich variety of 
highly energetic phenomena in and around clusters that seem capable of
energizing the electrons.
Still, today, the origin of cluster CRes is not
clear, although many proposals have been made.

%
%
%
%

%
\section{Cosmic Shock Waves} \label{shocks.se}
%
%
%
%
An early study of cosmic shocks was carried out in 
\citet{suze72s} who studied the evolution of individual 
perturbation modes till their nonlinear breakup, and \citet{bert85b}
who worked out self similar models describing the development of 
infall flows in an Einstein-de Sitter universe.
\citet{minetal00} have carried out a detailed study 
of cosmic shocks based on numerical simulations of structure formation.
Their analysis was aimed at establishing the statistical properties of 
cosmic shocks in view of their potential role as high energy particle 
accelerators. In fact, astrophysical shocks are collisionless and 
as part of the dissipation process generate, a supra-thermal 
distribution of high energy particles, namely cosmic-rays (CR).
%
   \begin{figure}
        \includegraphics[width=7cm]{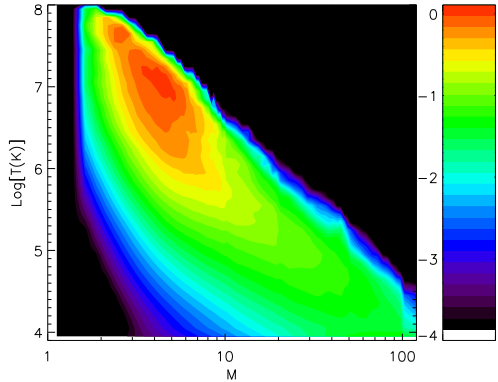}
      \caption{Top: Two dimensional diagram showing the differential 
of the thermal energy 
per unit log interval in both Mach number and temperature
produced at cosmic shocks throughout cosmic history.
It is shown in units of keV per particle.}
         \label{shfl2.fig}
   \end{figure}
%
Thus \citet{minetal00} for the first time plotted the distribution
of shocks as a function of Mach number. This is 
important to know for assessing the potential conversion of the
shocks kinetic energy (more precisely, ram pressure) into cosmic-rays. 
In Fig.~\ref{shfl2.fig} provides an example of such a diagram
showing, as a function of pre-shock temperature, 
$T_1$, and shock Mach number, $M$, the thermal energy  per unit log
$M$ and $T_1$ dissipated at cosmic shocks throughout cosmic history 
\citep[$\partial^2\Delta E_{th}/\partial\log M\partial \log T_1$][]{min02}.
The structure of the accretion shocks was found highly complex, irreducibly 
three dimensional and well beyond the capability of analytic description
\citep{minetal00}. Another important result, relevant for shock acceleration
(see below), was that a significant fraction 
of the shock energy is dissipated in relatively strong, high Mach
number shocks \citep{minetal00, min02}. The recent resolution study of 
\citet{rkhj03} confirms this earlier main findings of \citet{minetal00}.
They showed that as higher resolution is employed, weaker shocks become
more numerous, although the strong shocks remain unaffected.
The highest resolution simulation, with the same resolution as \citet{min02}
but twice as large a box, was characterised by 
a peak of the distribution plotted in Fig. \ref{shfl2.fig}
at slightly lower values than \citet{min02}.
While this does not affect the shocks mainly responsible for the high 
energy CRs, it is likely that most of the ``extra'' weak shocks 
in Ryu et al. are not 
even associated with virialized structures \citep{rkhj03}.

   \begin{figure}
        \includegraphics[width=7cm]{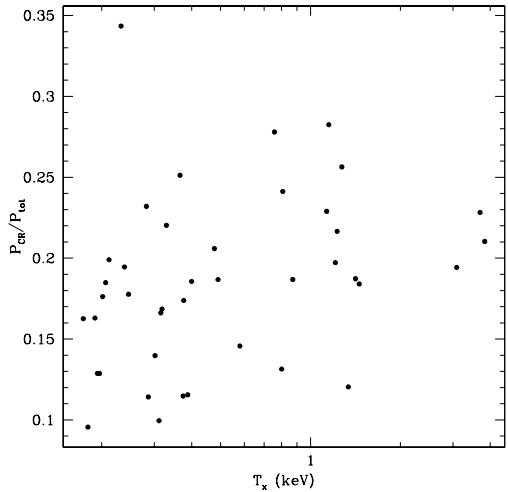}
      \caption{Ratio of CR to thermal pressure averaged over
the group/cluster volume within 1.0 h$^{-1}$ Mpc plotted as a function
of group/cluster core temperature.}
         \label{crp.fig}
   \end{figure}
%
\section{Cosmic-ray Pressure}

\cite{mrkj01} studied the 
production of CR protons at cosmic shocks by carrying out a numerical
simulation of structure formation that included {\it directly} shock
acceleration (in the test-particle limit approximation), transport and
energy losses of the CRs \citep{min01}.  CR injection takes place at
shocks according to the thermal leakage prescription, leading to the
injection, as CRs, of a fraction about $10^{-4}$ of the thermal protons
passing through a shock. According to those results, cosmic ray ions may
provide a significant fraction of the total pressure in the
intra-cluster medium (Fig. \ref{crp.fig}). This is basically because:
(i) intergalactic shocks are characterized by relatively large Mach
numbers \citep{minetal00,min02}; (ii) relativistic protons in the 
diffuse ICM are only
mildly affected by energy losses and (iii) up to energies $\sim
10^{15}$ eV are basically confined within Mpc scales  \citep{voahbr96}
by $\mu$G turbulent magnetic fields \citep{clkrbo00}.  
However, such conclusion
cannot be made strictly quantitative yet, primarily because it depends
on the unknown CR injection/acceleration efficiency at
shocks. Although a number of indirect argument can be made in order to
constrain the CR content in the ICM, such as radio emission from
secondaries \citep{blco99,mjkr01,min03}, the best observational
assessments will be possible through $\gamma$-ray emission (see
below).  In principle, though, CR pressure could account up to a few
tens of percents of the total ICM pressure. The cosmological
consequences of this result were discussed in \citet{mjkr01}.
\section{Radio Emission}

About 30\% of massive galaxy clusters (GC) exhibit diffuse radio 
synchrotron
emission extending over Mpc scales (Feretti, these proceedings).
Two broad classes have been identified.
{\it Radio halos} have a regular morphology resembling that
of the thermal X-ray emission
and no sign of polarization, whereas {\it radio relics} are
elongated, located at the periphery of a cluster,
and exhibit polarized emission (Feretti, these proceedings).
The short cooling time of the emitting CR electrons and the
large extension of the observed radio sources seem to require
ongoing acceleration mechanism in the intracluster medium (ICM).
\subsection{Radio Halos}
%
%
%
   \begin{figure}
      \caption{Diffuse synchrotron radio emission produced by shock accelerated electrons.}
         \label{radios.fig}
   \end{figure}
%
%
%
Radio halos are usually found in rich clusters with high ICM temperature,
$T\gsim 7$keV, and high X-ray luminosity, $L_x (0.1-2.4\mbox{keV})
\gsim 5\times 10^{44}$erg s$^{-1}$ (Feretti, these proceedings).
Since it usually extends over a linear size of about 1 \hinv Mpc,
the radio emission appears to be a characteristic of
the whole cluster, rather than being associated with any 
of the individual cluster galaxies \citep{willson70}. 
Signatures of a merging process in these clusters are often emphasized 
and the absence of cooling flows cited 
as demonstrating the connection between radio halos and mergers
\citep[e.g.][]{buote01}.
However, \citet{lhba00} \citep[also][]{bacchietal03} 
found a tight and steep correlation 
between the radio power emitted at 1.4 GHz and the cluster temperature,
and suggested that the apparent rarity of detections should be attributed 
to observational insensitivity to any but the most massive clusters.

A number of ideas have been proposed to explain the origin of the
relativistic electrons. It was soon realized that relativistic 
electrons ``ejected'' from cluster galaxies during a putative active phase
would {\it not} be able to reproduce the observed radio emission profile 
\citep[of Coma cluster; ][]{jaffe77,rephaeli77}. 
Alternative models commonly assume a continual energization of the
relativistic electrons by {\it in situ} first or second order Fermi 
processes.

Direct numerical simulations, however, show that shock accelerated
electrons produce radio maps where the emission has an irregular
morphology and is mostly concentrated in the outer regions
\citep[see Fig. \ref{radios.fig}; ][]{mjkr01}.
That is because in large virialized objects 
strong shocks occur at the outskirts. In any case this result
rules out shock acceleration mechanism for the origin of radio halos.

The radio electrons could also be produced as secondary products of
inelastic p-p collisions of CR ions and the thermal intra-cluster
nuclei \citep{dennison80,vestrand82,blco99,mjkr01,pfrommer04}.
\cite{mjkr01} computed self-consistently the population of both 
shock accelerated CR protons and the secondary e$^\pm$ they produce
in a simulation of structure formation and found that basically all 
general properties of radio halos (morphology, polarization, 
L$_{\rm radio}$ vs T$_x$ relation) where properly reproduced. 
However, they did predict a spectral steepening of the radio emission.
Yet, this does not exclude additional processes,
such as the interaction with weak and/or strong 
ICM turbulence, that would alter the CR protons distribution so
as to produce steepening features in the radio spectrum of the
secondary e$^\pm$. Additional arguments that disfavor secondary
models have been put forth by \cite{brunetti02} although these have 
been recently disputed by \cite{pfrommer04}.
In any case, the spectral steepening of Coma radio emission led 
\citet{schlick87} to consider in detail second order Fermi models 
where the particles are accelerated via their interactions with 
plasma waves generated by turbulence in the medium. These models 
are in general 
more flexible in that a number of free parameters in the transport 
equation of the CR electrons allows easier fit to the observations. 
There are, certainly, other more complex physical issues regarding the 
evolution of the turbulent spectrum, the generation of suitable waves 
for the acceleration process to properly work, the possible back 
reaction of the particles on the waves. For example,
very recently \citet{bbcg04} pointed out that the presence of
CR protons bearing just a few per cent of the ICM thermal pressure 
would damp most of the Alfv\`en waves, thus inhibiting turbulent 
acceleration of CR electrons. 
The most serious concern for Fermi-II mechanisms is due to
their inefficiency at accelerating the particles directly from the
thermal pool, a fact that was realized early on \citep{jaffe77}. It is
usually assumed, often rather casually, that the seed particles are
created at shock during merger events. Clearly modeling the CR particles
in the ICM is more
complex that initially anticipated and requires a combination of 
several processes including shock acceleration and turbulent 
reacceleration at some level. It is not clear, though,
what type of shock acceleration efficiencies are needed, both for 
of CR electrons and protons; if those requirements would lead to an 
overall consistent picture, that is if all other general
properties of radio halos would be properly reproduced. 
These issues will need to be addressed in the future.

\subsection{Radio Relics}
Clusters hosting radio relics
are somewhat less massive and cooler than those related 
to radio halos. They show no apparent correlation with
merger events, are observed in clusters
containing cooling flows \citep{bapili98} and
are found both near the cores of clusters and
at their outskirts. The spectra are typically steep, but 
explicit cutoffs are relatively rare even though the 
cooling time of the relativistic electrons is much shorter than 
the age of the cluster. 

The most likely mechanism for the origin of radio relic emission
is acceleration at accretion/merger shocks \citep{ebkk98,mjkr01}. 
In this respect, 
\citet{robust99} carried out simulation of binary merger and showed
that, as the main large 
scale shocks propagate out of the merging clusters, two post-shock 
arcs form which closely resemble the famous radio emission of A3667.
\cite{mjkr01} computed the evolution of shock accelerated CR electrons 
in a simulation of structure formation reproducing the main 
properties of radio emission, including radio power, morphology,
polarization and spectral index. 

Some of the peripheral radio emission, particularly those 
highly filamentary and extending over a few hundred kpc,
could be produced by an alternative mechanism, whereby
relic relativistic plasma previously injected by radio galaxies,
is being reenergized by the passage of an accretion/merger 
shock \citep{engo01}.

\section{EUV and HXR excesses}
In addition to radio emission, a number of clusters show emission at
extreme ultraviolet \citep[\eg][]{lieuetal96a} and hard x-rays
\citep[\eg][]{fufeetal99} in excess of what expected from the thermal
emission of the ICM.  Some of the physical implications associated
with these measurements were addressed in \citet[e.g.][]{mjkr01}.
Both measurements are extremely challenging and at the limit of the
instrumental capabilities. In fact, the observational results
concerning the EUV emission are still debated. Recently questions
have also been raised about the existence of HXR emission from Coma
cluster \citep{romo04}, although \citet{fufeetal04} confirm it.
%
%
%
   \begin{figure*}
   \centering
   \resizebox{\hsize}{!}{\includegraphics[clip=true]{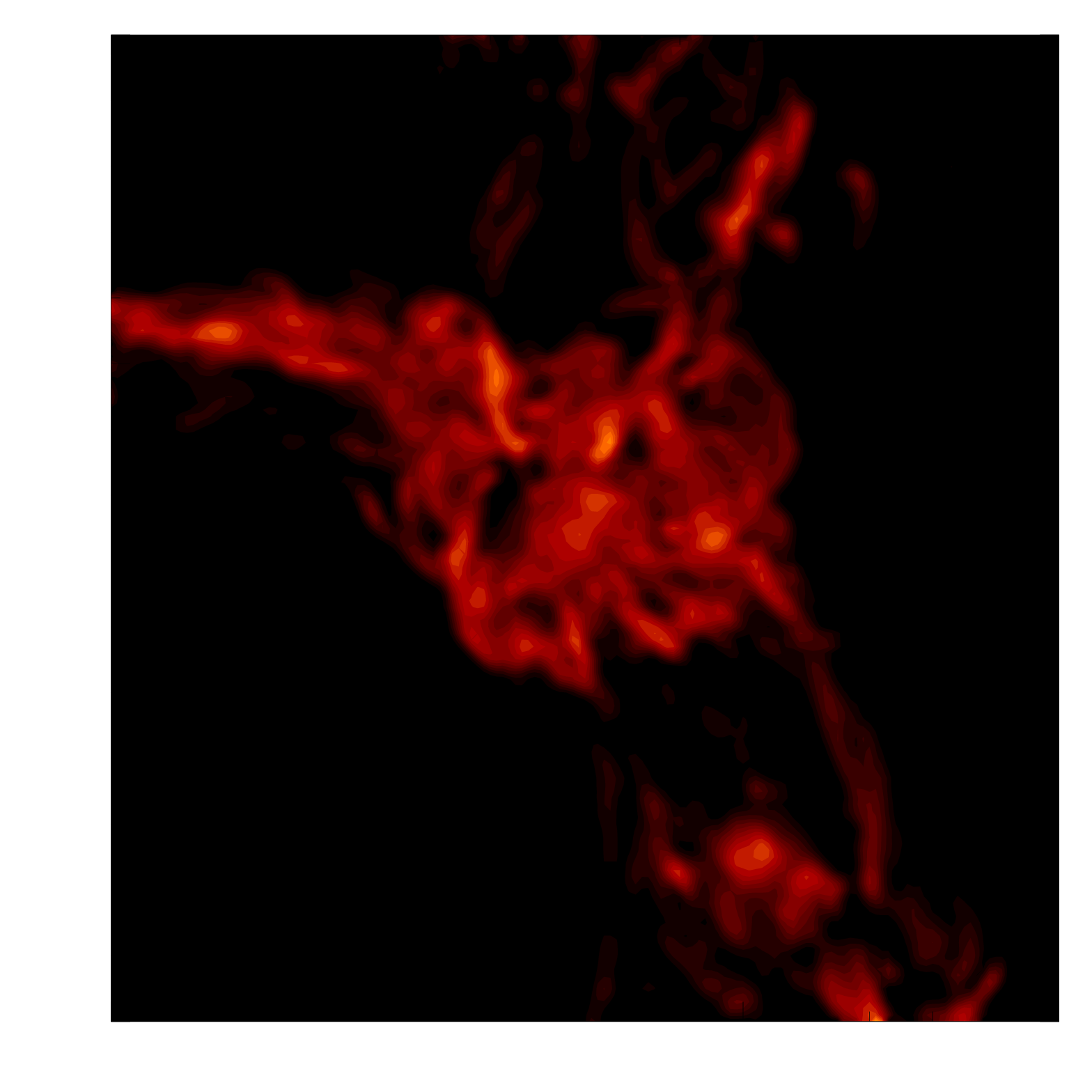}
        \includegraphics[width=24cm]{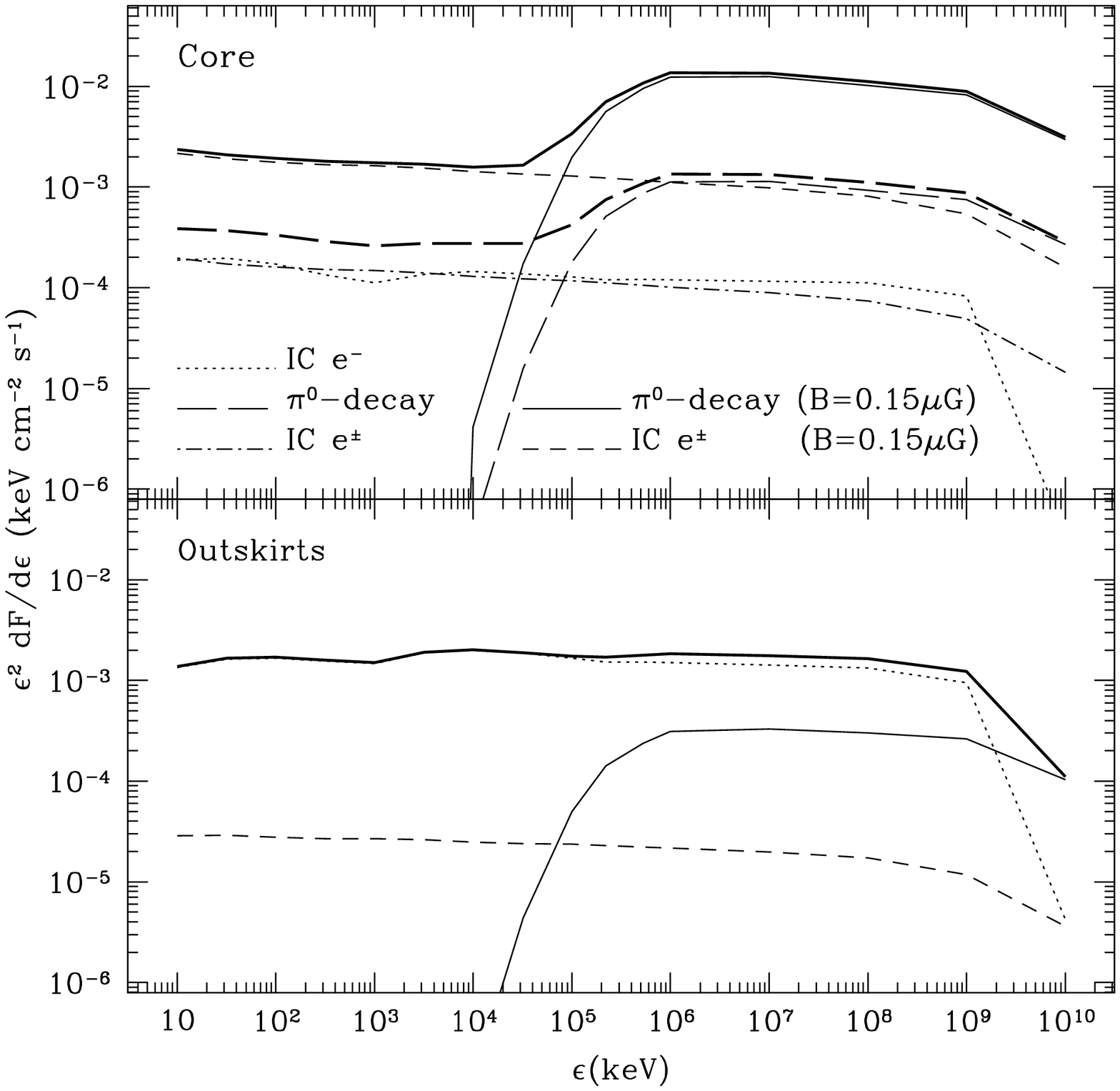}}
      \caption{ {\it Left}:
Synthetic map, 15 h$^{-1}$ Mpc on a side, of the integrated photon flux 
ranging from several $\times 10^{-9}$ to $10^{-11}$ in units 
``ph cm$^{-2}$ s$^{-1}$ arcmin$^{-2}$''. {\it Right}: synthetic spectra 
extracted from the inner region (top) and the outer region( (bottom). 
}
         \label{gam1.fig}
   \end{figure*}
%
%

\section{$\gamma$-ray Emission}
\gr observations of galaxy clusters are relevant for a number of 
reasons. They may provide direct evidence for the existence of 
CR protons in the ICM, which is important in order to determine 
the level of CR pressure there and whether or not secondary e$^\pm$ 
are viable models for radio halos. In addition, inverse Compton emission
from shock accelerated electrons is of great interest to determine
the contribution of this process to the \gr background and also 
to image and possibly investigate accretion shock physics \citep{min03}.
An important point is that, with conservative assumptions about the 
efficiency of acceleration of CR protons and electrons at cosmic shocks,
\gr fluxes due to IC and $\pi^0$-decay for a Coma-like cluster 
are expected to be comparable \citep[see Fig. \ref{gam2.fig}; ][]{min03}. 
In order for the observational results to be properly interpreted, one
should be able to discriminate between the two components.
As shown below, this should be possible for a nearby clusters, like Coma.

In the following I examine spectral and spatial properties
of \gr radiation between 10 keV and 10 TeV due to 
shock accelerated CRs in GCs. 
The relevant emission processes are: 
\pnd and IC emission from both shock accelerated (primary) 
CR electrons and secondary \epm 
(non-thermal bremsstrahlung turns out unimportant). 
The left panel in Fig. \ref{gam1.fig} shows 
a synthetic map of the integrated photon flux above 100 MeV
for a Coma like cluster of galaxies \citep{min03}.

Because of severe energy losses, 
\gr emitting primary electrons are only found 
in the vicinity of strong shocks where they are accelerated. 
Thus, the IC emission they produce is extended and reveals a rich
morphology reflecting the complex ``web'' of accretion 
shocks surrounding GCs \citep{minetal00}. 
On the other hand, the emission from \pnd and 
\epm is confined to the cluster core
where it creates a diffuse 
halo which rapidly fades with distance from the center.
In fact, \epm and $\pi^0$ are produced at the 
highest rate in the densest regions where both the parent CR ions 
and target nuclei are most numerous. 

These findings are further illustrated in the right panel of the 
same figure where synthetic spectra extracted from a core 
(top; with a $0.5^o$ radius) 
and an outskirts region (bottom; a ring with inner and outer 
radii of $0.5^o$ and $1.5^o$ respectively) are shown. 
As illustrated, the emitted radiation 
in the outskirts region is strongly 
dominated by \ic emission from primary e$^-$.
Conversely, in the core region $\pi^0$-decay (solid thin line)
dominates at high photon energy ($>$ 100 MeV) (top panel).
Notice that, given the observed radio flux for Coma cluster, 
if the assumption is made that the radio emission is produced
by secondary e$^\pm$, 
the total number of protons and of \epm 
depends on the magnetic field strength, $B$. Thus, two cases for
$\langle B \rangle$, namely 0.15 and 0.5 $\mu$G, are 
presented in the left panel of fig. \ref{gam1.fig}.
Both flux level from the inner and outskirts regions is above the
sensitivity limit of upcoming \gr facilities, particularly  GLAST and 
also Cherenkov telescopes \cite[MAGIC, HESS, VERITAS, 5@5; ][]{min03}. 
One delicate issue will
concern the performance of these telescope in the observation of 
extended sources, particularly for Cherenkov telescopes which have 
a relatively small field of view. Interestingly, the case for detection
of \gr emission from GCs has already been made in a number of cases
\citep{shmu02,pfrommer03,iudyn04}
although future experiments are strongly needed \citep{reposrma03}.

\section{Cosmic $\gamma$-ray Background}
\label{cgb.sec}
%
   \begin{figure*}
   \centering
   \resizebox{\hsize}{!}{\includegraphics[clip=true]{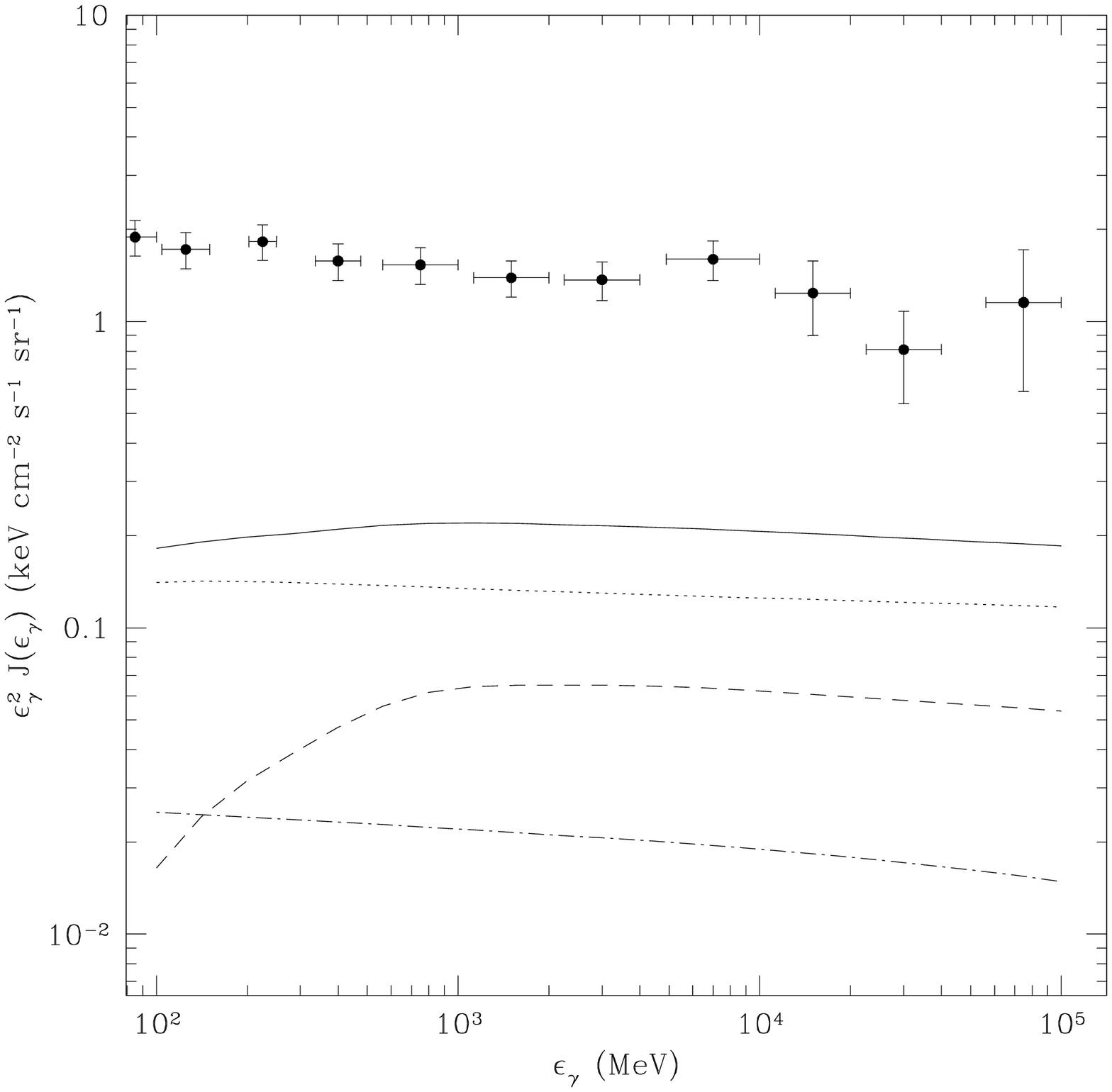}
        \includegraphics[clip=true]{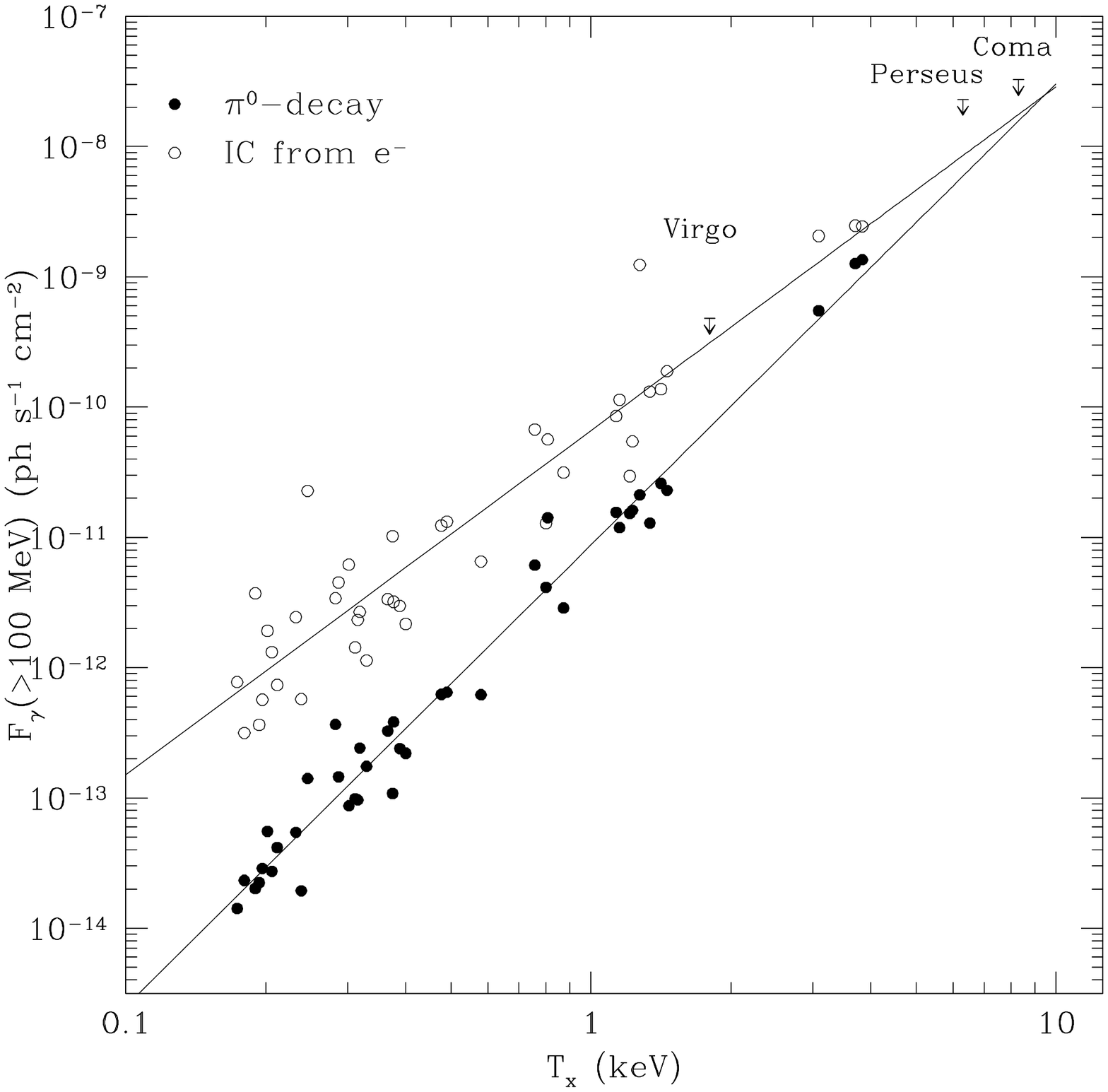}}
      \caption{ {\it Left}: \gr background: data (points with errorbars)
and model predictions (see text for details). {\it Right}: 
predicted cluster \gr luminosity versus cluster virial temperature.
EGRET upper limits are from \citet{reposrma03}.
}
         \label{gam2.fig}
   \end{figure*}
%
%
%
CR in the large scale structure may contribute 
a relevant fraction of the cosmic \gr background \citep[CBG; ][]{sreeku98}.
Although recently reassessed to half it initial value
\citep{stmore03}, much of the CGB is still 
unaccounted for \citep{chmu98}. 
Decay of neutral $\pi$-mesons can realistically
produce only a few percent of the measured CGB \citep{dasha95,cobl98};
however, IC emission from electrons accelerated at intergalactic shocks
is potentially more promising \citep{lowa00}.

Fig. \ref{gam2.fig} (left) shows the 
contribution to the CGB from CRs accelerated at cosmic
shocks according to a simulation of structure formation
that included the evolution of CR protons, electrons and 
secondary e$^\pm$.
EGRET observational data (solid dots) \citep{sreeku98}
are also shown for
comparison. Included are: IC emission 
of CR electrons scattering off cosmic microwave background 
photons (dot line), decay of neutral pions 
produced in p-p inelastic collisions (dash line) and
\ic emission from secondary \epm (dot-dash line). 
In Fig.~\ref{gam2.fig}, the total flux (solid line) 
corresponds roughly to a constant value at the level of 
0.2 keV cm$^{-2}$ s$^{-1}$ sr$^{-1}$ throughout the spectrum. 
It is dominated by \ic emission from primary 
electrons. Fractions of 
order 30\% and 10 \% are produced by $\pi^0$-decay 
and \ic emission from secondary e$^\pm$, respectively.
All three components produce the 
same flat spectrum, a reflection of the fact that the CRs
distributions were generated in strong shocks.
The computed flux is only $\sim$ 15 \% of the 
observed CGB by \citet{sreeku98}. It is difficult to imagine a higher 
contribution from \pnd and IC emission from 
\epm. In fact, if more CR protons were produced
at shocks, CR-induced shock modifications would actually
reduce the population of \gr emitting protons (and \epm).
On the other hand, the fraction, $\eta$, of shock ram 
pressure converted into 
CR electrons, can be constrained by
comparing  the simulated clusters' \gr photon luminosity
above 100 MeV to the upper limits set by the EGRET 
\citep{sreeku96,reposrma03} for nearby GCs.
This is done in Fig. \ref{gam2.fig} (right panel).
The simulation data (open circles) 
are best-fit by the curve (solid line):
\begin{eqnarray} \label{gamfit.eq}
{\rm L}_\gamma (>100~ {\rm MeV}) =
8.7\times 10^{43}  ~~~~~~~~~~ \nonumber\\  
\times \left(\frac{\eta}{4\times 10^{-3}}\right) 
\; \left(\frac{\rm T_x}{\rm keV}\right)^{2.6} \; {\rm ph ~s}^{-1}.
\end{eqnarray}
Thus, the EGRET upper limits require that 
$\eta \leq 0.8 \%$. This implies an upper limit on the computed 
\gr flux of about 0.35 keV cm$^{-2}$ s$^{-1}$ sr$^{-1}$
or a fraction of order $\sim 25$ \%  of the CGB.  
In view of the newly revised level of the CGB \citep{stmore03},
this contribution would fill the gap between the observed flux and 
that which is already accounted for.
%
%
%
%
%

\begin{acknowledgements}
I am grateful to the organizers for the hospitality and financial support
and to C. Pfrommer for reading the manuscript.
This work was partially supported by the Research
and Training Network `The Physics of the Intergalactic Medium',
EU contract HPRN-CT2000-00126 RG29185.
\end{acknowledgements}

\bibliographystyle{aa}
\bibliography{papers,books,papigm,proceed}

\begin{thebibliography}{44}
\expandafter\ifx\csname natexlab\endcsname\relax\def\natexlab#1{#1}\fi

\bibitem[{{Bacchi} {et~al.}(2003){Bacchi}, {Feretti}, {Giovannini}, \&
  {Govoni}}]{bacchietal03}
{Bacchi}, M., {Feretti}, L., {Giovannini}, G., \& {Govoni}, F. 2003, \aap, 400,
  465

\bibitem[{Bagchi {et~al.}(1998)Bagchi, Pislar, \& {Lima Neto}}]{bapili98}
Bagchi, J., Pislar, V., \& {Lima Neto}, G.~B. 1998, \mnrasl, 296, L23

\bibitem[{Bertschinger(1985)}]{bert85b}
Bertschinger, E. 1985, \apjs, 58, 39

\bibitem[{Blasi \& Colafrancesco(1999)}]{blco99}
Blasi, P. \& Colafrancesco, S. 1999, \app, 12, 169

\bibitem[{Brunetti(2002)}]{brunetti02}
Brunetti, G. 2002, in Matter and Energy in Clusters of Galaxies, ed. S.Bowyer
  \& C.-Y. Hwang, ASP, Taiwan, astro-ph/0208074

\bibitem[{{Brunetti} {et~al.}(2003){Brunetti}, {Blasi}, {Cassano}, \&
  {Gabici}}]{bbcg04}
{Brunetti}, G., {Blasi}, P., {Cassano}, R., \& {Gabici}, S. 2003, ArXiv
  Astrophysics e-prints

\bibitem[{{Buote}(2001)}]{buote01}
{Buote}, D.~A. 2001, \apjl, 553, L15

\bibitem[{Chiang \& Mukherjee(1998)}]{chmu98}
Chiang, J. \& Mukherjee, R. 1998, \apj, 496, 752

\bibitem[{Clarke {et~al.}(2001)Clarke, Kronberg, \& B\"ohringer}]{clkrbo00}
Clarke, T.~E., Kronberg, P.~P., \& B\"ohringer, H. 2001, \apjl, 547, L111

\bibitem[{Colafrancesco \& Blasi(1998)}]{cobl98}
Colafrancesco, S. \& Blasi, P. 1998, \app, 9, 227

\bibitem[{Dar \& Shaviv(1995)}]{dasha95}
Dar, A. \& Shaviv, N.~J. 1995, \prl, 75, 3052

\bibitem[{Dennison(1980)}]{dennison80}
Dennison, B. 1980, \apjl, 239, L93

\bibitem[{En{\ss}lin {et~al.}(1998)En{\ss}lin, Biermann, Klein, \&
  Kohle}]{ebkk98}
En{\ss}lin, T.~A., Biermann, P.~L., Klein, U., \& Kohle, S. 1998, \aap, 332,
  395

\bibitem[{En{\ss}lin \& Gopal-Krisna(2001)}]{engo01}
En{\ss}lin, T.~A. \& Gopal-Krisna. 2001, \aap, 366, 26

\bibitem[{Fusco-Femiano {et~al.}(1999)Fusco-Femiano, {Dal Fiume}, Feretti,
  Giovannini, Grandi, Matt, Molendi, \& Santangelo}]{fufeetal99}
Fusco-Femiano, R., {Dal Fiume}, D., Feretti, L., {et~al.} 1999, \apjl, 513, L21

\bibitem[{Fusco-Femiano {et~al.}(2004)Fusco-Femiano, Orlandini, Brunetti,
  Feretti, Giovannini, Grandi, \& Setti}]{fufeetal04}
Fusco-Femiano, R., Orlandini, M., Brunetti, G., {et~al.} 2004, \apjl

\bibitem[{{Iyudin} {et~al.}(2004){Iyudin}, {B{\" o}hringer}, {Dogiel}, \&
  {Morfill}}]{iudyn04}
{Iyudin}, A.~F., {B{\" o}hringer}, H., {Dogiel}, V., \& {Morfill}, G. 2004,
  \aap, 413, 817

\bibitem[{Jaffe(1977)}]{jaffe77}
Jaffe, W.~J. 1977, \apj, 212, 1

\bibitem[{Leslie \& Elsmore(1961)}]{leel61}
Leslie, P. R.~R. \& Elsmore, B. 1961, The Observatory, 81, 14

\bibitem[{Liang {et~al.}(2000)Liang, Hunstead, Birkinshaw, \&
  Andreani}]{lhba00}
Liang, H., Hunstead, R.~W., Birkinshaw, M., \& Andreani, P. 2000, \apj, 544,
  686

\bibitem[{Lieu {et~al.}(1996)Lieu, Mittaz, Bowyer, Lockman, Hwang, \&
  Schmitt}]{lieuetal96a}
Lieu, R., Mittaz, J. P.~D., Bowyer, S., {et~al.} 1996, \apjl, 458, L5

\bibitem[{Loeb \& Waxman(2000)}]{lowa00}
Loeb, A. \& Waxman, E. 2000, \nat, 405, 156

\bibitem[{Miniati(2001)}]{min01}
Miniati, F. 2001, \cpc, 141, 17

\bibitem[{Miniati(2002)}]{min02}
Miniati, F. 2002, \mnras, 337, 199

\bibitem[{Miniati(2003)}]{min03}
Miniati, F. 2003, \mnras, 342, 1009

\bibitem[{Miniati {et~al.}(2001{\natexlab{a}})Miniati, Jones, Kang, \&
  Ryu}]{mjkr01}
Miniati, F., Jones, T.~W., Kang, H., \& Ryu, D. 2001{\natexlab{a}}, \apj, 562,
  233

\bibitem[{Miniati {et~al.}(2001{\natexlab{b}})Miniati, Ryu, Kang, \&
  Jones}]{mrkj01}
Miniati, F., Ryu, D., Kang, H., \& Jones, T.~W. 2001{\natexlab{b}}, \apj, 559,
  59

\bibitem[{Miniati {et~al.}(2000)Miniati, Ryu, Kang, Jones, Cen, \&
  Ostriker}]{minetal00}
Miniati, F., Ryu, D., Kang, H., {et~al.} 2000, \apj, 542, 608

\bibitem[{{Pfrommer} \& {En{\ss}lin}(2003)}]{pfrommer03}
{Pfrommer}, C. \& {En{\ss}lin}, T.~A. 2003, \aap, 407, L73

\bibitem[{{Pfrommer} \& {En{\ss}lin}(2004)}]{pfrommer04}
{Pfrommer}, C. \& {En{\ss}lin}, T.~A. 2004, \aap, 413, 17

\bibitem[{Reimer {et~al.}(2003)Reimer, Pohl, Sreekumar, \& Mattox}]{reposrma03}
Reimer, O., Pohl, M., Sreekumar, P., \& Mattox, J.~R. 2003, \apj, 588, 155

\bibitem[{Rephaeli(1977)}]{rephaeli77}
Rephaeli, Y. 1977, \apj, 212, 608

\bibitem[{Roettiger {et~al.}(1999)Roettiger, Burns, \& Stone}]{robust99}
Roettiger, K., Burns, J.~O., \& Stone, J.~M. 1999, \apj, 518, 603

\bibitem[{Rossetti \& Molendi(2004)}]{romo04}
Rossetti, M. \& Molendi, S. 2004, \aap, astro-ph/0312447

\bibitem[{Ryu {et~al.}(2003)Ryu, Kang, Hallman, \& Jones}]{rkhj03}
Ryu, D., Kang, H., Hallman, E., \& Jones, T.~W. 2003, \apj, 593, 599

\bibitem[{{Scharf} \& {Mukherjee}(2002)}]{shmu02}
{Scharf}, C.~A. \& {Mukherjee}, R. 2002, \apj, 580, 154

\bibitem[{Schlickeiser {et~al.}(1987)Schlickeiser, Sievers, \&
  Thiemann}]{schlick87}
Schlickeiser, R., Sievers, A., \& Thiemann, H. 1987, \aap, 182, 21

\bibitem[{Sreekumar {et~al.}(1996)Sreekumar, Bertsch, Dingus, Esposito,
  Fichtel, Fierro, Hartman, Hunter, Kanbach, Kniffen, Lin, Mayer-Hasselwander,
  Mattox, Michelson, von Montigny, Mukherjee, Nolan, Schneid, Thompson, \&
  Willis}]{sreeku96}
Sreekumar, P., Bertsch, D.~L., Dingus, B.~L., {et~al.} 1996, \apj, 464, 628

\bibitem[{Sreekumar {et~al.}(1998)Sreekumar, Bertsch, Dingus, Esposito,
  Fichtel, Hartman, Hunter, Kanbach, Kniffen, Lin, Mayer-Hasselwander,
  Michelson, von Montigny, Muecke, Mukherjee, Nolan, Pohl, Reimer, Schneid,
  Stacy, \& et~al.}]{sreeku98}
Sreekumar, P., Bertsch, D.~L., Dingus, B.~L., {et~al.} 1998, \apj, 494, 523

\bibitem[{Strong {et~al.}(2003)Strong, Moskalenko, \& Reimer}]{stmore03}
Strong, A.~W., Moskalenko, I.~V., \& Reimer, O. 2003, in , ICRC No.~28 (UAP),
  astro-ph/0306345

\bibitem[{Sunyaev \& Zel'dovich(1972)}]{suze72s}
Sunyaev, R.~A. \& Zel'dovich, Y.~B. 1972, \aap, 20, 189

\bibitem[{Vestrand(1982)}]{vestrand82}
Vestrand, W.~T. 1982, \aj, 87, 1266

\bibitem[{V\"olk {et~al.}(1996)V\"olk, Aharonian, \& Breitschwerdt}]{voahbr96}
V\"olk, H.~J., Aharonian, F.~A., \& Breitschwerdt, D. 1996, \ssr, 75, 279

\bibitem[{Willson(1970)}]{willson70}
Willson, M. A.~G. 1970, \mnras, 151, 1

\end{thebibliography}

\end{document}